\documentclass[twocolumn,twocolappendix]{aastex7}

\usepackage{enumitem}
\usepackage{float}
\usepackage{array,booktabs,makecell}
\setlist[itemize]{noitemsep}
\graphicspath{{./plots/}}
\usepackage{graphicx}
\usepackage{changepage} 
\usepackage[export]{adjustbox}
\usepackage{threeparttable}

\newcommand{\zogy}{\texttt{ZOGY} } 

\newcolumntype{M}{>{\texttt}l}

\makeatletter
\let\frontmatter@title@above=\relax
\makeatother

\begin{document}

\title{Ghosts of eruptions past: Searching for historical Galactic supernovae using variable thermal dust echoes and machine learning}

\author[0000-0003-1481-8076]{Justin Vega}
\affiliation{Department of Astronomy, Columbia University, 550 W 120th Street, New York NY 10027, USA}
\email{j.vega@columbia.edu}

\author[0000-0003-1481-8076]{Kishalay De}
\affiliation{Department of Astronomy, Columbia University, 550 W 120th Street, New York NY 10027, USA}
\affiliation{Center for Computational Astrophysics, Flatiron Institute, 162 Fifth Avenue, New York, NY 10010, USA}
\email{kde3038@columbia.edu}

\author[0000-0003-2242-0244]{Ashish Mahabal}
\affiliation{Division of Physics, Mathematics and Astronomy, California Institute of Technology, Pasadena, CA 91125, USA}
\affiliation{Center for Data Driven Discovery, California Institute of Technology, Pasadena, CA 91125, USA}
\email{aam@astro.caltech.edu}

\author[]{Jacob E. Jencson}
\affiliation{IPAC, Mail Code 100-22, Caltech, 1200 E. California Boulevard, Pasadena, CA 91125, USA}
\email{}

\author[0000-0003-2758-159X]{Viraj R. Karambelkar}
\altaffiliation{NASA Hubble Fellow}
\affiliation{Department of Astronomy, Columbia University, 550 W 120th Street, New York NY 10027, USA}
\email{}

\author[]{Armin Rest}
\affiliation{Space Telescope Science Institute, 3700 San Martin Dr., Baltimore, MD 21218, USA}
\affiliation{Physics and Astronomy Department, Johns Hopkins University, Baltimore, MD 21218, USA}
\email{}

\author[0000-0003-4127-0739]{Megan Masterson}
\affiliation{MIT Kavli Institute for Astrophysics and Space Research, Massachusetts Institute of Technology, Cambridge, MA 02139, USA}
\email{}

\shorttitle{Variable dust echoes in NEOWISE}
\shortauthors{Vega et al.}

\correspondingauthor{Kishalay De}
\email{kd3038@columbia.edu}

\begin{abstract}
The Galactic core-collapse supernova (SN) rate is estimated at $\approx 1–3$ per century; however, no optically visible SN has been discovered in the past 400 years. Although records of the last optically detected SN (Cassiopeia A) are debated, it is revealed today via its bright, variable mid-infrared (MIR) dust echoes -- offering the possibility of identifying dust-obscured, missed events via their dust echoes. We present the first all-sky, untargeted search for thermal dust echoes of luminous Galactic transients using difference imaging on 12 years of time-resolved NEOWISE co-adds (spanning $2009-2022$) followed by statistical detection of variable extended sources. We use echo features around Cas A, together with archival catalogs to train a convolutional neural network to classify transient candidates as dust echoes, point sources, artifacts, and high proper motion stars. Our model achieves $\approx 94$\% accuracy in distinguishing echoes from other variable sources. Applying the classifier to $\approx 11$ million transient candidates, we search for spatial over-densities of echoes across the Galactic plane. We find that Cas A is the only region exhibiting echoes at the WISE sensitivity threshold of $W2$ surface brightness of $\approx 20$\,Vega\,mag\,arcsec$^{-2}$ -- reflecting its unique combination of young age and luminous shock breakout. We present the largest catalog of time-resolved echo positions of Cas A (20,477 within 10$^\circ$) that are being used for studies of the surrounding interstellar medium with the {\it James Webb Space Telescope}. Our results lay the groundwork for the imminent {\it Roman} space telescope surveys -- which will achieve $\approx 100\times$ higher sensitivity and $\approx 30\times$ better spatial resolution at wavelengths of $\lesssim 2.5\,\mu$m.
\end{abstract}


\section{Introduction} \label{sec:intro}

Estimates for the rate of Galactic core-collapse supernovae (SNe) range between $\approx 1-3$ events per century \citep{Adams_2013, Rozwadowska_2021}, although the last convincingly observed optical Galactic SN dates back to the Kepler's SN in 1604 \citep{Murphey2021}. There is evidence for more recent Galactic SNe, including the Cassiopeia A (Cas A) and the G1.9+0.3 which was discovered as a radio SN remnant dating back to $\approx 100$\,yrs \citep{Green_2008, Luken2020}. While optical observations certainly suffer from severe dust extinction in discovering Galactic SNe \citep{Adams_2013}, the current gap of $\gtrsim 400$\,yr may soon become alarmingly long given increasing synoptic survey capabilities for bright IR transients \citep{De2020, Dong2025, Kasliwal2025}. Not only would a new Galactic SN offer exceptional insights into the (poorly understood) SN mechanism, even identification of a previously missed, recent remnant would offer new clues to SN ejecta properties given the proximity of Galactic events \citep{Rest2011, JWSTCasA}.

The last (possibly) optically detected Galactic SN was Cas A. While there is contested evidence of its direct observation $\approx 400$ years ago \citep{CasA_history, Hughes_1980}, the light from its explosion is observed today through interactions with dust in the surrounding interstellar medium. Scattered light echoes from dust surrounding Cas A have been observed at optical wavelengths \citep{casa_lightecho}, allowing a precise measurement of its type \citep{CasAIIb}, distance \citep{Neumann2024} and geometry \citep{Rest2011}. However, the SN light is also absorbed and re-radiated by surrounding dust clouds, producing thermal dust echoes \citep{irecho_theory} observed in mid-infrared (MIR) light that reveal its luminous shock breakout \citep{Dwek2008} and its interaction with the surrounding medium \citep{Krause2005, CasA_ISM, Vogt_2012, JWSTCasA}.

Importantly, the bright MIR echoes of Cas A are easily identifiable in time-resolved MIR images \citep{Dwek2008, Besel_2012}, offering the possibility of identifying historically missed luminous Galactic transients by a systematic search for MIR echoes. This provides a novel, complementary way of discovering recent SNe compared to SN remnant searches -- which are limited by dust extinction in optical/X-ray bands, and spatial resolution and low surface brightness features in the radio bands \citep{Green2025}. While scattered light optical echoes and dust echoes have been previously discovered around known SN remnants \citep{Krause2005, Rest2005}, untargeted searches require synoptic surveys as well as systematic classification techniques to remove the large foreground of other variable sources.

Synoptic repeated mapping of the IR sky has become possible only in recent years. The WISE mission \citep{Wright2010}, reinitiated in 2014 as the NEOWISE survey \citep{Mainzer2014} has carried out the longest ($\approx 15$\,year) and only full-sky survey in the MIR $W1$ ($3.4\,\mu$m) and $W2$ ($4.6\,\mu$m) bands with a cadence of six months, offering the unique possibility to search for variable MIR echoes of recent Galactic SNe that may have been missed. Although the survey was originally designed for solar-system objects, recent works have employed image subtraction techniques on archival NEOWISE data to systematically search for transient events. In this paper, we present the first systematic search for dust echoes in all-sky, time resolved difference images from the NEOWISE mission. We employ recent advances in large-scale applications of deep learning based image classification using convolutional neural networks (CNNs) in wide-field imaging surveys. In Section \ref{sec:data}, we describe the NEOWISE data, the pipeline to detect and manage transient candidates, and the pre-processing of our training set for machine learning applications. Section \ref{sec:model} details our CNN model and the training and validation process. In Section \ref{sec:results}, we perform classification for all of our transient candidates and present general results from our analysis. We summarize our results and future outlook in Section \ref{sec:discuss}.

\section{Data}\label{sec:data}

\begin{figure*}[!ht]
    \includegraphics[width=\textwidth]{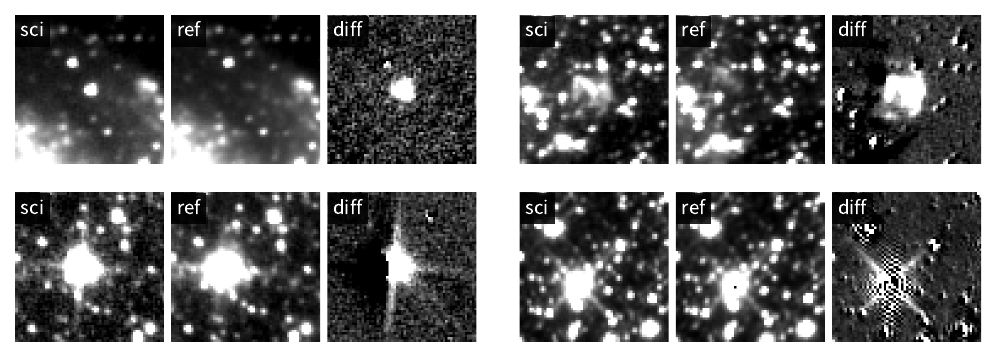}
    \caption{Examples of the four classes of transient candidates identified in our classification architecture. We show cutout triplets of the science (single epoch image), reference (template image), and difference images from left to right for a point source transient (top left), a characteristic yin-yang pattern for a high proper motion star (bottom left), a dust echo (top right) and an artifact (bottom right). Cutouts are North-up, East-left and at the native \texttt{unwise} pixel scale ($2.75$"/pixel). Note that visible residuals in the static background sources depend on both the image scaling (scaled between $-1\sigma$ and $+3\sigma$ of the cutout, where $\sigma$ is the cutout standard deviation) and the expected source (Poisson) noise \citep{zogy}; therefore the visible residuals are more significant near brighter background sources and where the overall cutout variance is smaller.}
\label{fig:cutouts}
\end{figure*}

\subsection{Transient candidate pipeline}\label{sec:pipeline}
In this work, we use a custom image subtraction pipeline from \citet{De2020} based on the \zogy algorithm \citep{zogy}. We perform image subtraction on the NEOWISE single-epoch co-added images in $W1$ and $W2$ bands from the \texttt{unwise} project \citep{UnWISEUNBLURREDCOADDS2014, FULLDEPTHCOADDSWISE2017}, using the full-depth co-added image from the original WISE survey and the first NEOWISE epoch (between $\approx 2009-2014$) as the reference image of a given field. When including the pause in WISE observations between 2011 - 2014, the reference image effectively includes four full passes of the sky over a baseline of $\approx 5$\,years. While the temporal baseline does smear out variability in some sources, our choice to use this stack as the reference image is motivated by the substantial improvement in image quality due to mitigation of artifacts \citep{FULLDEPTHCOADDSWISE2017} including the NEOWISE data.

From the \zogy difference images, we filter for candidates with statistical significance of $\geq 7\sigma$ \citep{zogy}. For each transient that passed this threshold, we create cropped 61$\times$61 pixel cutout triplets from the science, reference, and difference images, and store the cutouts in a PostgreSQL database. For our transient search, we use all NEOWISE data obtained between 2014 and 2023 (i.e. 
excluding the reference image epochs), corresponding to 18 epochs of observations over most of the sky\footnote{At the time of model creation, complete \texttt{unwise} images were available for this time range since the last two NEOWISE epochs were released later. Since echoes are persistent variable sources, we do not expect the use of 18 epochs to affect our search technique.}. The sample of transient candidates has been previously used to carry out wide searches for point sources characteristics of compact eruptive transients (e.g. \citealt{Zuckerman2023, masterson2024, Myers2024}). However, large excess residuals from known dust echoes are also flagged in this method -- as the light echo front propagates through the interstellar medium, the regions of dust emission are seen to move across the sky \citep{Krause2005}. Therefore subtraction of time-resolved images against a constant template produce (often bright) extended residuals.

\subsection{Training Sample}
We begin by constructing a training sample of transient types for our machine learning application using CNNs. Since the image of statistical significance in \zogy (called Scorr) is agnostic to the exact nature of the residual\footnote{We note here that the $7\sigma$ threshold used in selecting candidates only correctly applies to point sources, since the \zogy method is optimized for point source recovery. However, sufficiently bright extended sources are also flagged, and we attempt to quantify the corresponding threshold in Section \ref{sec:results}.}, the transient candidates also include non-astrophysical artifacts (e.g. diffraction spikes from bright stars), subtraction residuals from moving objects (e.g. high proper motion stars) and residuals from extended variable emission (e.g. dust echoes). While this work is focused on the unique capabilities of this dataset to search for dust echoes as extended variable sources, we choose to create a multi-class classifier that sorts the transient candidates into four broad groups -- point source transients (hereafter `real'), extended variable emission (hereafter `echo'), artifacts from optical elements (hereafter `artifact') and high proper motion stars (hereafter `highpm'). The scheme makes the classifier even more broadly useful for separating point source transients from other contaminants.

Figure \ref{fig:cutouts} shows examples of science, reference and difference cutout images for each of these classes. Transient candidates were primarily labeled using a combination of both human classification and cross-matching to known variable sources. To build the sample of reals, we supplemented human-labeled candidates by cross-matching the database with spectroscopically confirmed extragalactic transients in the Transient Name Server\footnote{https://wis-tns.org}. For proper motion stars, we supplement the human-labeled sample with candidates coincident with Gaia DR3 stars \citep{GAIA_dr3} with measured proper motions $> 100$\,mas/year ($> 0.3$ pixel offsets in $\approx 10$\,yrs). We performed targeted searches around the Cas A region to find and visually label candidates consistent with dust echoes. In training our model using echoes around Cas A, we expect it to be representative of any variable extended features surrounding Galactic transients -- given the morphological similarities of these features to the extensive samples of scattered light echoes of Galactic transients \citep{Rest2005, casa_lightecho, Rest2012} . All candidates used for training were visually verified prior to model implementation. Our training set includes 7,200 unique candidates across 18 epochs. The number of candidates in each class are equally distributed by design to mitigate the effects of class imbalance on model training.

\subsection{Pre-processing}

For our deep learning classification, we found that the differentiation between these classes of subtraction residuals was more effective (both visually and for our model) when analyzing all epochs of data for a given candidate, i.e. using a `movie' of the variable emission over all the epochs of NEOWISE data. An example of a full set of difference images for an echo candidate can be seen in Figure \ref{fig:model-arch}. While previous works of wide-field surveys have primarily used triplet cutouts (Figure \ref{fig:cutouts}) for training and evaluation of deep learning applications \citep{braii, De2020, BTSbot}, we found a significant increase in classification accuracy for echoes when using the difference images (only) across all epochs. We provide further details of these tests and model performance in Appendix \ref{sec:app}. However, we also developed a model that functions with single epoch triplets to aid in point source recovery applications for transient searches; we describe this model in Appendix \ref{sec:singleepoch}.

We train, validate and evaluate our model on cutout arrays centered at the position of each transient -- comprising of 18 epochs of NEOWISE difference images available across most of the sky\footnote{We exclude images at very high ecliptic latitudes $|\beta| > 80^\circ$, since \texttt{unwise} images in these regions are often incomplete due to the NEOWISE scanning strategy \citep{FULLDEPTHCOADDSWISE2017}. However, this region does not include any portion of the Galactic plane.}. Prior to the CNN implementation, we standardized and normalized the data to ensure uniformity across all images. To standardize our image data, we follow \citet{braii} and clip the pixel values of each individual cutout image to fall between $3\sigma$ of the median pixel value, with $\sigma$ representing the standard deviation of the pixel values of the image array. We also normalize each cutout to values between 0 and 1. This procedure ensures that the model is agnostic to overly bright/dim pixel values present in the images, which can influence the feature extraction process in deep learning models. The resulting unit-normalized $61 \times 61 \times 18$ array is then used for training and validation; the same technique is used for inferencing on the sample of unseen candidates.

\section{Model}\label{sec:model}

\begin{figure*}[t!]
    \centering
\includegraphics[width=\linewidth]{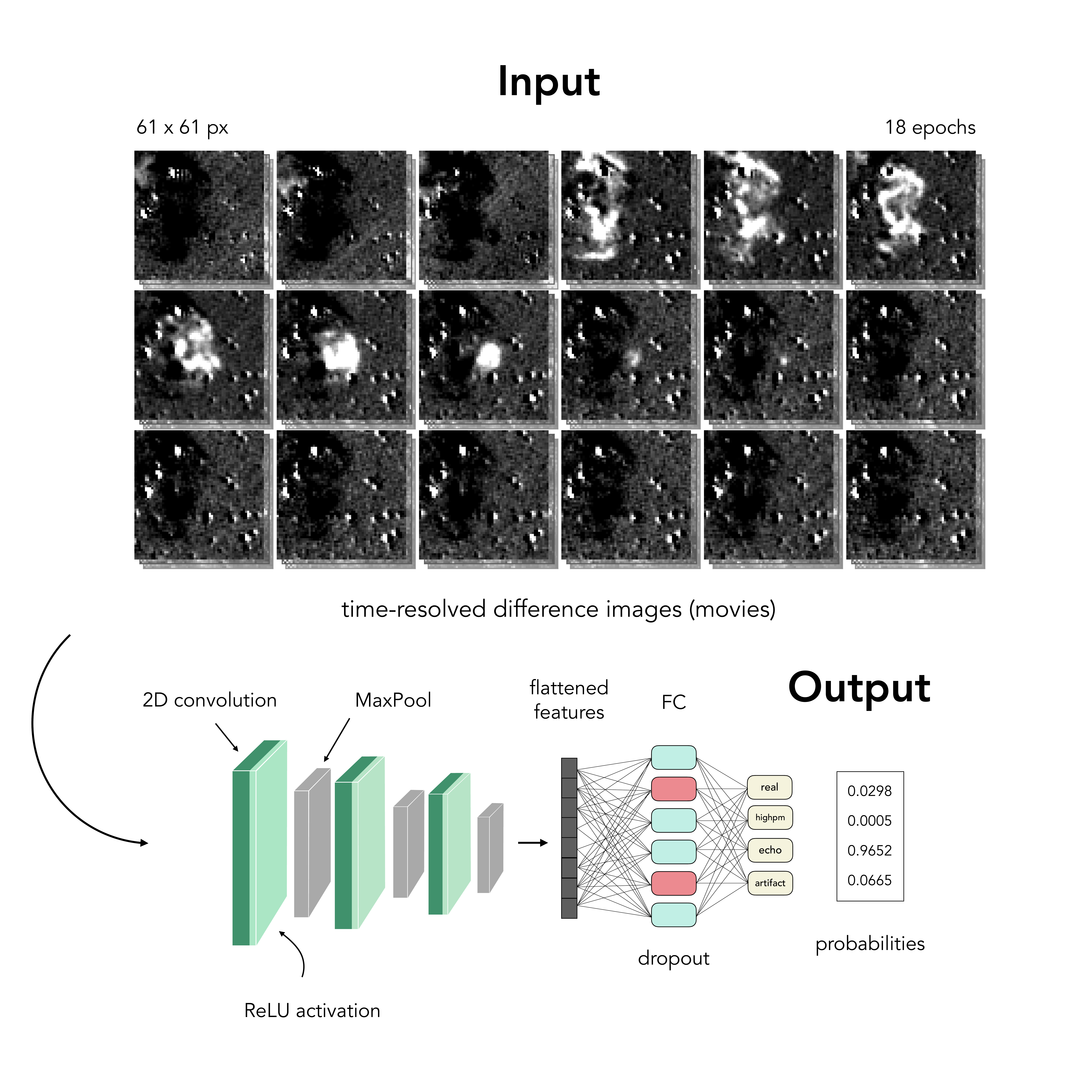}
    \caption{The model architecture of the CNN. The epochal time-resolved difference images for a transient candidate are arranged as a 61 x 61 x 18 array as an input for the model; the example shown are the difference images for an echo location near Cas A with an extended structure seen propagating from left to right in the difference images. ReLU activation is used for the convolutional and dense layers, and softmax activation is used for the final layer to map class probabilities between 0 and 1.}
    \label{fig:model-arch}
\end{figure*}

To construct the convolutional neural network, we used the \texttt{Keras} API within \texttt{TensorFlow} \citep{tensorflow, keras}. We use the sparse categorical cross-entropy loss function and Adaptive Moment Estimation (Adam) optimizer \citep{ADAM}, using a batch size of 32. The data are equally distributed between classes into 80\% training, 10\% testing, and 10\% validation to avoid data leakage during model development. We adopt a modified version of the VGG architecture, as originally proposed by \citet{vgg}. The CNN architecture and pipeline is shown in Figure \ref{fig:model-arch}. Broadly, features of the input data are learned by downsampling the data via kernel convolution operations. Each convolutional layer contains kernels (filters) that extract features from the input images, and an activation function that allows the model to learn non-linear features. The resulting feature maps are then downsampled via a pooling layer. With each successive map, the model learns more distinct features (i.e. edges to shapes), and these features are condensed for classification.

We construct a classification model with three convolutional layers with a kernel size of $3 \times 3$ and Rectified Linear Unit (ReLU) activation, each followed by max pooling layers of size $2 \times 2$ (\texttt{MaxPool}). Learned features are then passed through a two-layer neural network, where the final (dense) layer is a linear layer with softmax activation that assigns probabilities for class membership. 

\subsection{Training}

We tune the model hyperparameters that control the architecture (e.g. the number of units of each convolutional/neural network layer, activation functions) and training process (e.g. number of epochs, learning rate, batch size) in accordance with loss and accuracy curves to achieve optimal model performance. To determine the optimal set of hyperparameters, we performed Bayesian hyperparameter sweeps using the Weights \& Biases API\footnote{\url{https://wandb.ai}}, assigning a range of options for each hyperparameter. A table of hyperparameters and their values are provided in Table \ref{tab:multi_hps} of Appendix \ref{sec:app}.

For all model runs, we also implement several forms of implicit and explicit regularization to prevent the model from overfitting. An early stopping criterion ensures that training does not continue if the model does not improve over a number of epochs, and only the best model is saved (we use 10 epochs; see Appendix \ref{sec:app}). A dropout layer was added to prevent units in the dense layer from explicitly ``memorizing'' feature patterns specific to the training set. We also augment the number of training samples by adding orthogonal rotations and horizontal and vertical flips of each series of cutouts to the training set, effectively increasing our number of samples for each class by a factor of 6. Our final choice in hyperparameters were chosen by selecting a set that maximizes the training and validation accuracy and minimizes training and validation loss. The loss and accuracy curves of the final model are shown in Figure \ref{fig:lossaccuracy} of Appendix \ref{sec:app}.

\subsection{Validation}\label{sec:val}

\begin{figure*}[t!]
    \includegraphics[width = \linewidth]{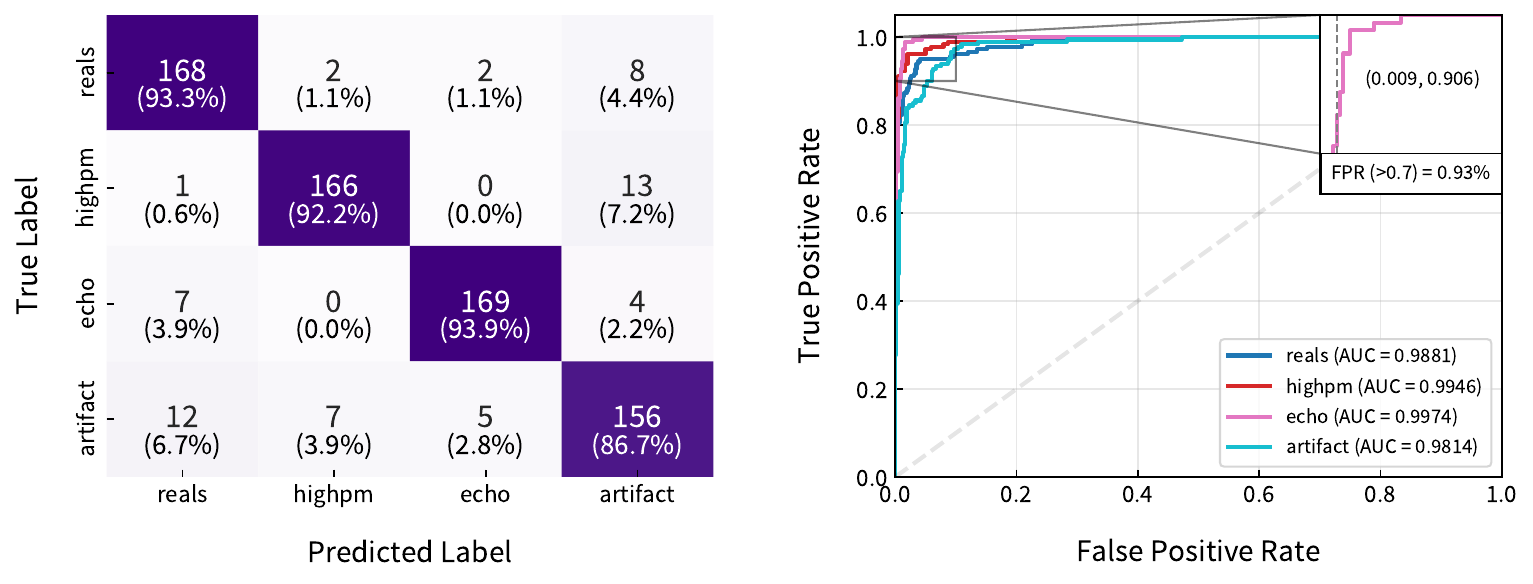}
    \caption{(Left) The confusion matrix for the best performing model, as calculated using a test set of 720 candidates. The distribution of examples for each class are equal across classes. The percentages represent the fraction over the total number of sources for a particular class. (Right) The ROC curve for each class in the model, characterizing each as a one-vs-all binary classification. We determine a threshold of 0.7 for true positives of echoes at a FPR of 0.93\%. The inset shows a zoom-in of the top-left edge of the curve, used to define the score threshold. A perfect classifier would have a TPR of 100\% at all thresholds, while a classifier that assigns classes at random would match the diagonal dashed gray line.}
    \label{fig:confusion}
\end{figure*}

Our classifier returns predicted class probabilities in the range [0, 1]. For each source, we choose the class with the highest probability (score) as the predicted class. The resulting confusion matrix as evaluated on the test set is shown in Figure \ref{fig:confusion}. From evaluating on a test set of 720 candidates, the model has a balanced accuracy and recall of 91.5\%, and a precision of 91.6\%. Echo candidates in particular have a true positive rate (TPR) of 93.9\%, which shows a strong indication of an acceptably accurate classifier. However, the confusion matrix indicates that the model has some difficulty with correctly identifying artifacts, as they have the lowest TPR of 86.7\%. 

We also evaluate the accuracy of the classifier by examining the true positive rate (TPR) against the false positive rate (FPR) at different thresholds, using the Receiver Operating Characteristic (ROC) curve. The area under the ROC curve (AUC) is also used as a summary statistic for the performance of a classifier. We adapt the metric for multi-class classification and show the ROC curve in Figure \ref{fig:confusion}. For all classes, the ROC curve has a strong elbow shape in the top left corner, indicating a well-performing classifier. The ROC curve for echoes performs the best out of the classes with an AUC = 0.997, followed by highpm, reals, and artifacts. In order to optimize detection of echoes for this work, we use the ROC curve -- we determine a probability threshold (score) of 0.7 for echoes at a FPR of 0.93\%, which we adopt for evaluating on unseen candidates in Section \ref{sec:results}. We make the production multi-epoch echo model and single epoch model publicly available on GitHub\footnote{\url{https://github.com/vegajustin26/neowise-ML}}, and include a tutorial on basic usage with example data. The complete list of echo positions will be made available in electronic format along with this publication. 

\section{Search for dust echoes} \label{sec:results}

\begin{figure*}[t!]
    \centering
    \includegraphics[width = \linewidth, scale = 0.95]{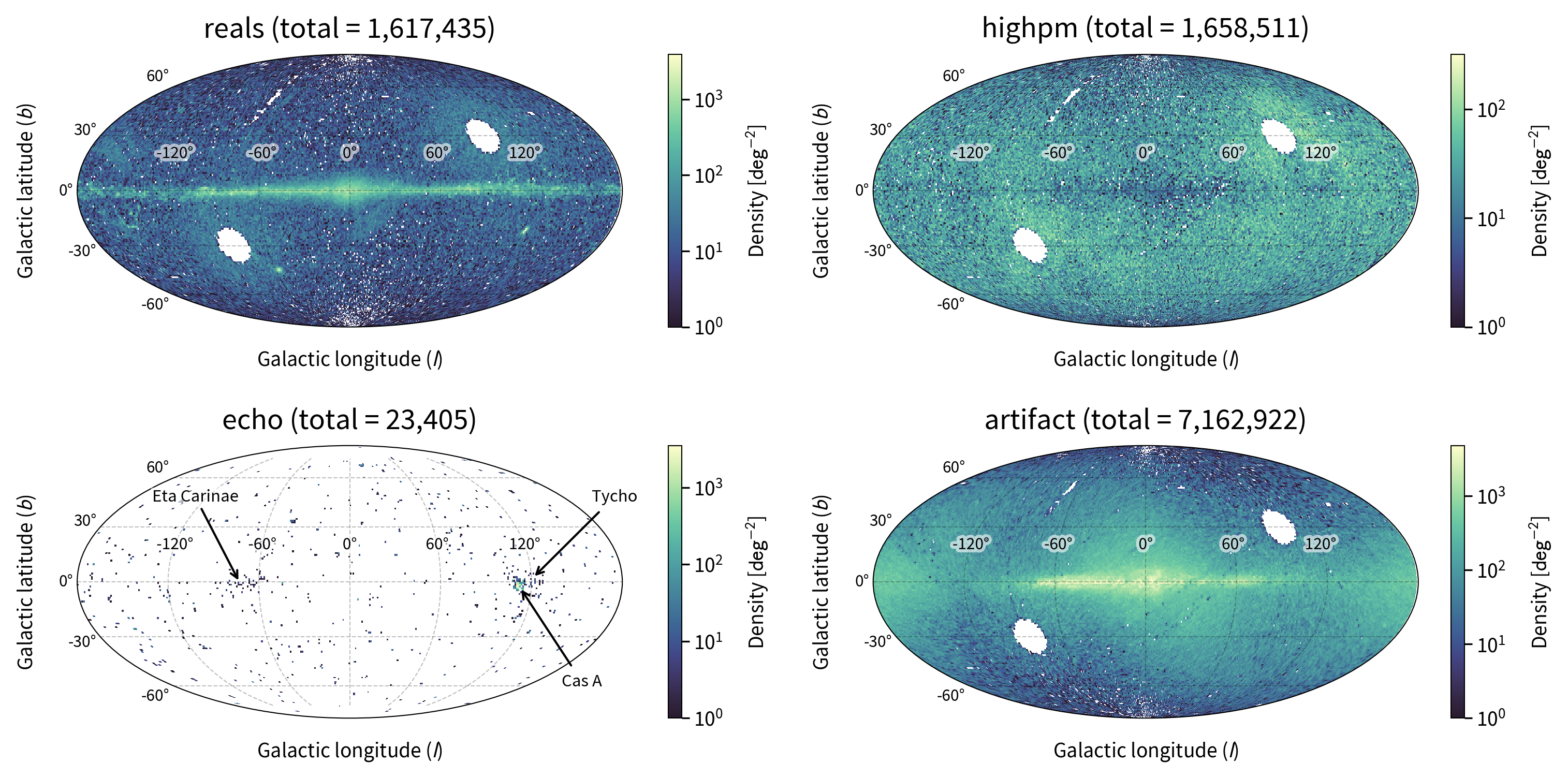}
    \caption{The sky density distribution of all transient candidates from each predicted class, with the total classified number indicated in the title. The location of the Cas A remnant is shown in the bottom left panel, together with other Galactic transients where scattered light echoes have been reported (Eta Carinae and Tycho's remnant). The white circular gaps correspond to regions at high ecliptic latitude ($|\beta| > 80^\circ$) with incomplete image coverage, that were excluded from the search.}
    \label{fig:skydist}
\end{figure*}

Our CNN model was designed to carry out an untargeted search for variable dust echoes in all of NEOWISE data. We run the classification model on a complete set of 11.1 million transient candidates -- representing all statistically significant positive residuals\footnote{Since negative residuals are naturally produced as a result of echo propagation (Figure \ref{fig:model-arch}), we do not separately classify or evaluate those candidates.} detected in difference images (Section \ref{sec:data}). We show a sky distribution of the predicted classes for the classifications in Figure \ref{fig:skydist}. The sky distribution for classified reals is concentrated near the Galactic plane as expected for stellar sources, while sources outside the plane are dominated by Active Galactic Nuclei \citep{Assef2018}. On the other hand, proper motion stars have a much more uniform distribution across the sky, reflecting the expected population of nearby stellar sources with measurable proper motion over the $\approx 10$\,year baseline \citep{Marocco2021}. The large number of bright IR sources in WISE results in artifacts being the most common type of subtraction residual, with a sky distribution tracking that of reals. This is expected since the sky distribution of bright stars also traces the Galactic plane.

\begin{figure*}[!ht]
    \centering
    \includegraphics[width=\linewidth]{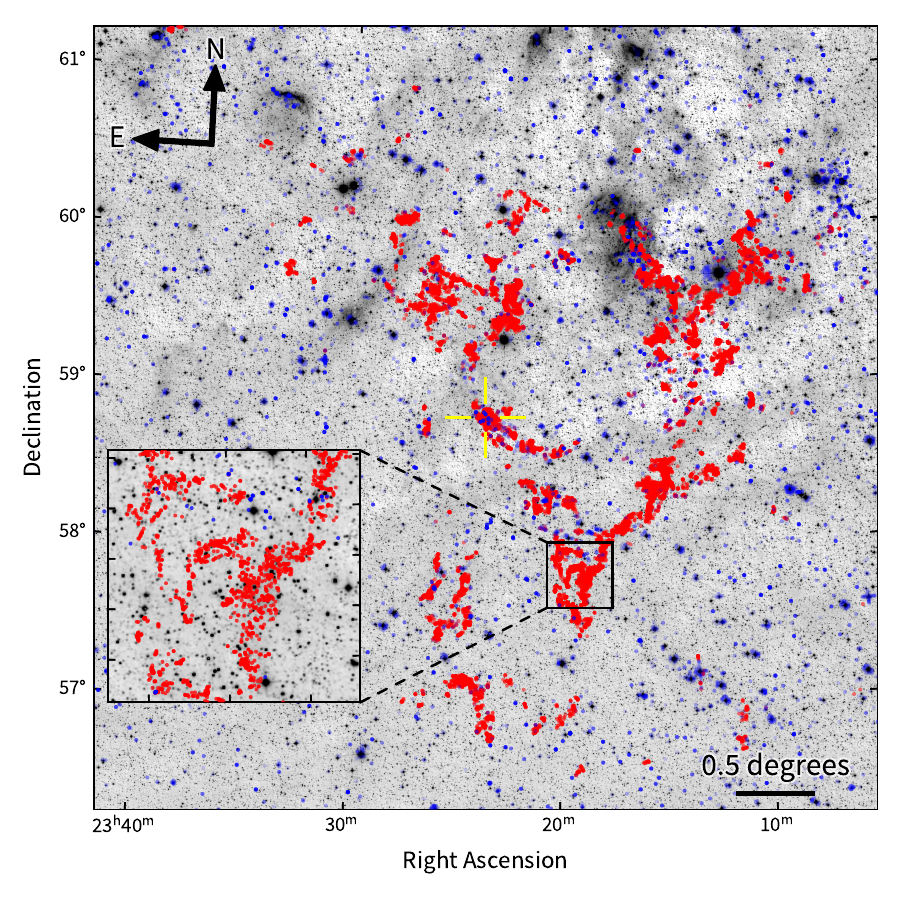}
    \caption{Co-added $5^\circ \times 5^\circ$ $W2$-band image (between $2009-2022$) of the region surrounding Cas A taken from \texttt{unwise}, overlaid with a map of the echo positions (red dots) and all other sources (blue dots) identified in our search. The black square inset shows a zoom-in of the region with a particularly high spatial density complex of echo features. The position of Cas A is indicated by the yellow cross-hair, and the image spatial scale and orientation are indicated.}
    \label{fig:casa}
\end{figure*}

\begin{figure*}[!ht]
\includegraphics[width=\linewidth]{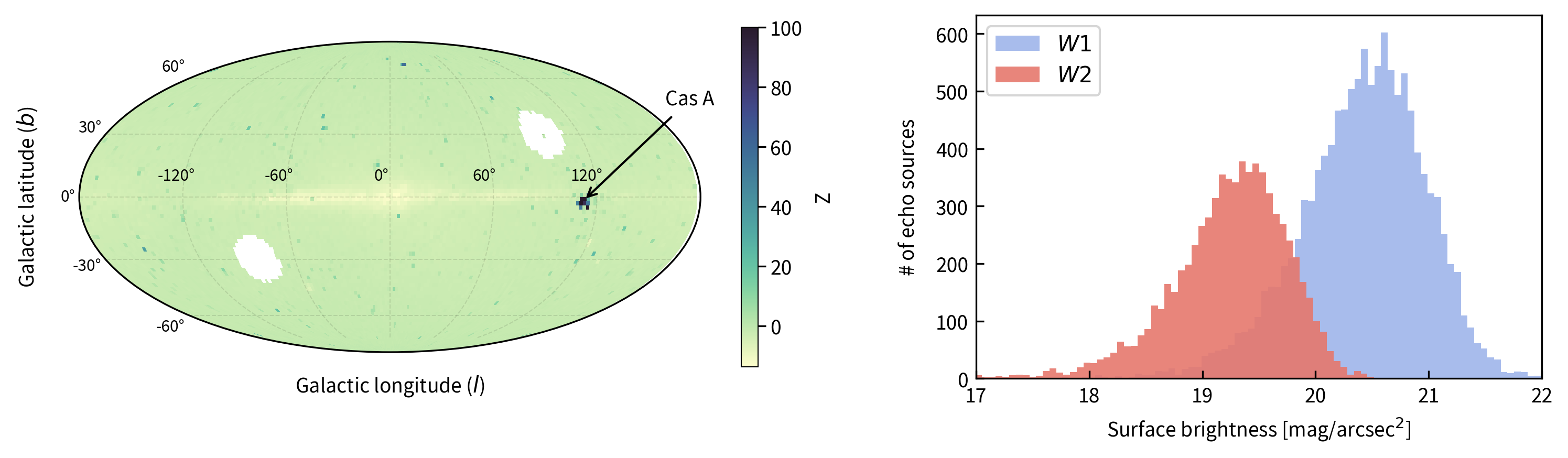}
    \caption{({\it Left}) All-sky map of the test-statistic $Z$ (color bar; see text) quantifying the statistical excess of echo candidates in each sky pixel of $\approx 2$\,sq.\,deg. The location of the Cas A remnant is marked. ({\it Right}) Histogram of the estimated surface brightness of echo candidates within $\approx 3^\circ$ of Cas A (Figure \ref{fig:casa}) in the $W1$ and $W2$ bands. The distributions drop off sharply at $\approx 21$\,Vega\,mag\,arcsec$^{-2}$ and $\approx 20$\,Vega\,mag\,arcsec$^{-2}$ in the $W1$ and $W2$ bands respectively.}
    \label{fig:zmap}
\end{figure*}

In total, the model identified 27373 possible echo candidates (i.e. with echo score exceeding that of all the other classes), including 23405 found at a score $> 0.7$ (Figure \ref{fig:echohist}). Figure \ref{fig:skydist} shows a large excess of echo candidates coincident with the known dust echoes surrounding Cas A remnant, in addition to other sources distributed along the Galactic plane. These echoes were first identified by \citet{Krause2005} with two epochs of pointed observations from the Spitzer space telescope. \citet{CasA_ISM} subsequently carried out pointed time series observations of the dust echoes using Spitzer, identifying a large number of variable extended structures. The all-sky capabilities of NEOWISE allow us to re-create the largest map of current echo positions, identified out to $\approx 7-8^\circ$ from the remnant. In particular, the total sample of identified echoes exceeds our training set by $\gtrsim 10\times$, suggesting that the model is able to reliably identify echo features spanning a wide range of background environments (both near and far from the remnant; Figure \ref{fig:casa}); however, it is difficult to independently quantify its efficacy in the absence of other known echo regions.

Figure \ref{fig:casa} shows the positions of identified variable echo locations along with other non-echo variables, overlaid on a static $W2$ image centered at the Cas A remnant. Although the central remnant is well detected in the co-added image, the propagating echoes are smoothed out in co-addition; however, our image differencing method clearly brings out smooth structures reflecting the location of surrounding dust clouds \citep{CasA_ISM}. Figure \ref{fig:casa} clearly demonstrates that our classification scheme is effective at separating the connected web of echo features from unrelated variables/artifacts among the transient candidates flagged in this region. The two dimensional echo structures in Figure \ref{fig:casa} are very similar to that reported by \citet{CasA_ISM} using pointed Spitzer observations in $2005$ over a smaller sky footprint.

Next, we turn to the search for echoes from previously unknown remnants. Given the very large number of transient candidates near the Galactic plane, we expect a substantial number of candidates to be incorrectly classified as echoes (given the estimated FPR $\approx 1-5$\%; Figure \ref{fig:confusion}, right panel). Figure \ref{fig:skydist} also shows the positions of other Galactic transients with previously identified scattered light echoes in optical light. As regions surrounding luminous Galactic transients are expected to host a large number of spatially adjacent echo features (as in Cas A; Figure \ref{fig:casa}), we attempt to search for sky regions with substantial excess of classified echoes relative to the estimated number from our FPR $f$. If $N_t$ is the total number of candidates in a sky direction and $N_e$ is the number classified as echoes, with the null hypothesis being of no excess of echo-like candidates, we can write the test statistic $Z$ as
\begin{equation}
    Z = \frac{N_{e} - f \times N_t}{\sqrt{f \times N_t \times (1 - f)}}
\end{equation}
where the denominator is the expected variance for a binomial distribution.

Figure \ref{fig:zmap} shows a sky projection of the test-statistic $Z$ for the sample of echo candidates, binned into sky regions of $\approx 2$ square degrees. We caution that $Z$ is useful to detect outliers only in the limit of large $N_t$, which is primarily valid near the Galactic plane where a large number of candidates are detected (Figure \ref{fig:skydist}). Since we are interested in searching echo positions from historically missed events near the Galactic plane, this limitation does not affect our analysis. Examining both the two-dimensional sky distribution (Figure \ref{fig:zmap}) and corresponding one-dimensional distribution within $|b| < 10^\circ$, we find that Cas A is the only clear region with statistical excess of $Z > 5$ near the Galactic plane\footnote{We visually examined the echo candidates close to Eta Carinae (Figure \ref{fig:skydist}) and found them to be consistent with consistent with false positives from optical artifacts of bright stars. While some echoes were identified spatially adjacent to Tycho's remnant, we visually confirmed that their propagation directions were consistent with the most distant echoes ($\approx 8^\circ$ away) from the Cas A remnant. Our visual examination of the remaining candidates away from this region suggest that they arise from mis-classification of other source types (particularly artifacts) by the model. The number of classified echoes ($\sim {\rm few} \times 10^3$) outside ($>10^\circ$ away) of the Cas\,A remnant (Figure \ref{fig:skydist}) is nominally consistent with the expected FPR ($\lesssim 1$\%; Figure \ref{fig:confusion}) when evaluated on $\sim 10^6$ unique sky positions of the complete candidate set.}.

To quantify the sensitivity of our search, Figure \ref{fig:zmap} also shows the distribution of estimated surface brightnesses of the echo candidates identified within $\approx 2^\circ$ of Cas A. For simplicity, the surface brightnesses were measured by performing aperture photometry at the location of peak echo flux (without adopting an extended structure model) and dividing it by the effective areal footprint of the aperture. The analysis shows that the faintest echo features are measured at a depth of $\approx 21$\,Vega\,mag\,arcsec$^{-2}$ and $\approx 20$\,Vega\,mag\,arcsec$^{-2}$ in the $W1$ and $W2$ bands respectively, providing an estimate of the depth of our search. For comparison, the single epoch point source depth of the \texttt{unwise} images is $W2 \approx 16.1$\,Vega\,mag \citep{Meisner2023} -- which approximately represents the threshold from our adopted significance cut for point sources in the \texttt{ZOGY} difference images. Scaling the limiting point source flux (distributed over the effective area of $\approx 25$\,arcsec$^2$ for the \texttt{unwise} point spread function in $W2$; \citealt{Lang2014}) to the corresponding limiting surface brightness yields a very similar estimate of $\approx 19.8$\,Vega\,mag\,arcsec$^{-2}$ for the expected sensitivity of the search.

\section{Summary and Future Prospects}
\label{sec:discuss}

We have presented a novel approach to identifying echoes of historically missed Galactic SNe via searches for their variable dust echoes. We leverage the unique wavelength coverage and time baseline of the NEOWISE mission together with image differencing techniques and modern machine learning techniques (with CNNs) to carry out an all-sky search for MIR echoes. The architecture classifies subtraction residuals in differencing into four classes (Figure \ref{fig:cutouts}) -- point sources, high proper motion stars, artifacts and extended echoes, using cutout movies of time-resolved NEOWISE images (Figure \ref{fig:model-arch}) over a 12-year baseline. Our model achieves a 94\% accuracy in identifying echo candidates with a FPR of 1\% (Figure \ref{fig:confusion}). We also develop model that functions on single epoch cutouts geared for point source transient identification in Appendix \ref{sec:singleepoch}. By evaluating the model on a sample of $\approx 11$M transient candidates identified from NEOWISE difference imaging, we examine locations of the known Cas A echoes as well as carry out untargeted searches.

We find a total of 20,477 variable echo positions within $10^\circ$ of the Cas A remnant. These echoes trace the structure of surrounding dust clouds (Figure \ref{fig:casa}), providing a map of present-day echo locations. These positions are serving as a critical resource for planning observations of the echoes with the {\it James Webb Space Telescope} \citep{Jencson2024, Peek2025} to study the effect of SN light on the surrounding medium, and using it as a beacon to map its fine structure. Given the lack of alternative wide-area IR survey capabilities, modeling the temporal propagation of the positions over the NEOWISE baseline will also offer the potential for future observations. We search for previously unknown echoes of luminous transients by evaluating statistical excesses of echo features across the sky. We do not find any additional accumulations of echo features down to our estimated sensitivity threshold of $\approx 20$\,AB\,mag\,arcsec$^{-2}$ in the $W2$ band. Given that multiple Galactic SNe are known from around the time of the Cas A explosion, the non-detection of additional clumps of echoing clouds suggests a unique combination of circumstances for Cas A -- while its young age and relative proximity sustain spatially well-resolved echoes, the luminous shock breakout of the Type IIb supernova \citep{CasAIIb} in ultraviolet light \citep{Dwek2008} provides an efficient dust heating mechanism that is not expected from remnants like Tycho's supernova (which occurred spatially and temporally close to Cas A but has been identified as a Type Ia supernova; \citealt{Krause2008}).

While the Cas A dust echoes today span a size of $\approx 3^\circ$ on the sky, our generalized search for echoes would have been sensitive to substantially smaller echoing regions (i.e. the projected size of the sky region where an echo excess may be present) given the WISE spatial resolution ($\approx 2.75$"/pixel), cutout sizes employed ($\approx 2.8$') and our adopted pixelation of the echo search ($\approx 1$\,sq.\,deg.; Figure \ref{fig:skydist}). Ultimately, our search is limited by the raw sensitivity of the WISE mission (which limits the detection of fainter echoes) and its spatial resolution (which limits the detection of echoes with slower propagation) -- soon to be surpassed by the {\it Roman Space Telescope} surveys.

With its $\approx 100\times$ higher sensitivity ($\approx 24$\,AB\,mag for point sources against $\approx 19$\,AB\,mag for WISE) and $\approx 30\times$ better spatial resolution ($\approx 0.2$\arcsec vs. $\approx 6$\arcsec for WISE), {\it Roman} will be exquisitely sensitive to echoes of both luminous and faint transients. Specifically, the High-Latitude Wide Area Survey will image dozens of nearby galaxies over two epochs \citep{Sanderson2026} in the same filter, offering the potential to directly apply similar techniques for temporally variable extended sources. Within the galaxy, the {\it Roman} Galactic Plane Survey (RGPS; \citealt{RGPS}) will map $\approx 1000$\,sq.\,deg. of the central Galactic Plane between 2027 - 2029, including time domain surveys of an area of $\approx 20$\,sq.\,deg.

In particular, given that the expected Galactic core-collapse SN (which should produce luminous shock break-out emission) rate is expected to be $\gtrsim 1/100$\,yr$^{-1}$, our WISE null detection of new echo regions suggest that the most recent Galactic SNe (as well as the youngest known remnant G1.9+0.3) may be located far into the Galactic disk where the majority of the Galactic star formation lies \citep{Bronfman2000}. While the extremely high surface brightness of this region as well as the expected small echo region size precludes a sensitive search with the low spatial-resolution WISE data, the RGPS will enable exquisite searches for these events employing similar techniques. However, we highlight two major challenges for the RGPS: i) dust emission is much fainter at shorter wavelengths, reducing the sensitivity of RGPS operating at $\lambda \lesssim 2.5\,\mu$m) and ii) the planned survey largely involves single visits of fields in a given filter; multi-epoch measurements may be possible by combining observations across multiple filters, requiring the inclusion of chromatic dust properties for optimal detection and modeling.

\software{
    \texttt{TensorFlow} \citep{tensorflow}; \texttt{Keras} \citep{keras}; \zogy \citep{zogy};
    \texttt{astropy} \citep{astropy:2013, astropy:2018, astropy:2022}; 
    \texttt{matplotlib} \citep{hunter2007matplotlib};
    \texttt{numpy} \citep{oliphant2006guide, walt2011numpy, harris2020array}; \texttt{scipy} \citep{scipy}; \texttt{tqdm} \citep{tqdm}; \texttt{q3c} \citep{Koposov2019}
}

\section*{Acknowledgements}

The authors are indebted to the Backyard Worlds team for making animations of the WISE/NEOWISE data publicly available\footnote{\url{http://byw.tools/wiseview}}. This expedited the data labeling process significantly. We acknowledge the support of the National Aeronautics and Space Administration through ADAP grant number 80NSSC24K0663 and Roman WFS grant number 80NSSC26K0113. The computations reported in this paper were (in part) performed using resources made available by the Flatiron Institute. The Flatiron Institute is funded by the Simons Foundation. The authors graciously thank Tiffany Suen for help with labeling transient candidates. We also thank Aaron Meisner for making the NEOWISE image co-adds accessible for analysis. J.V. acknowledges support from the National Science Foundation Graduate Research Fellowship under grant DGE-2437839.


\appendix

\section{Echo model development and performance}
\label{sec:app}
We provide additional details about the development and performance of our echo classification model. Our preliminary echo model was originally developed using triplet cutouts as our input data. However, candidates with characteristic time-varying effects (e.g. echo propagation, optical `ghosting' artifacts occurring every other epoch) were commonly misclassified in our single-epoch model, which elucidated the need for the model to learn from multi-epoch data. We find that training on multi-epoch difference images provided a notable increase in accuracy compared to triplet cutouts.

\begin{figure}[h]
    \centering
    \includegraphics[width = \linewidth]{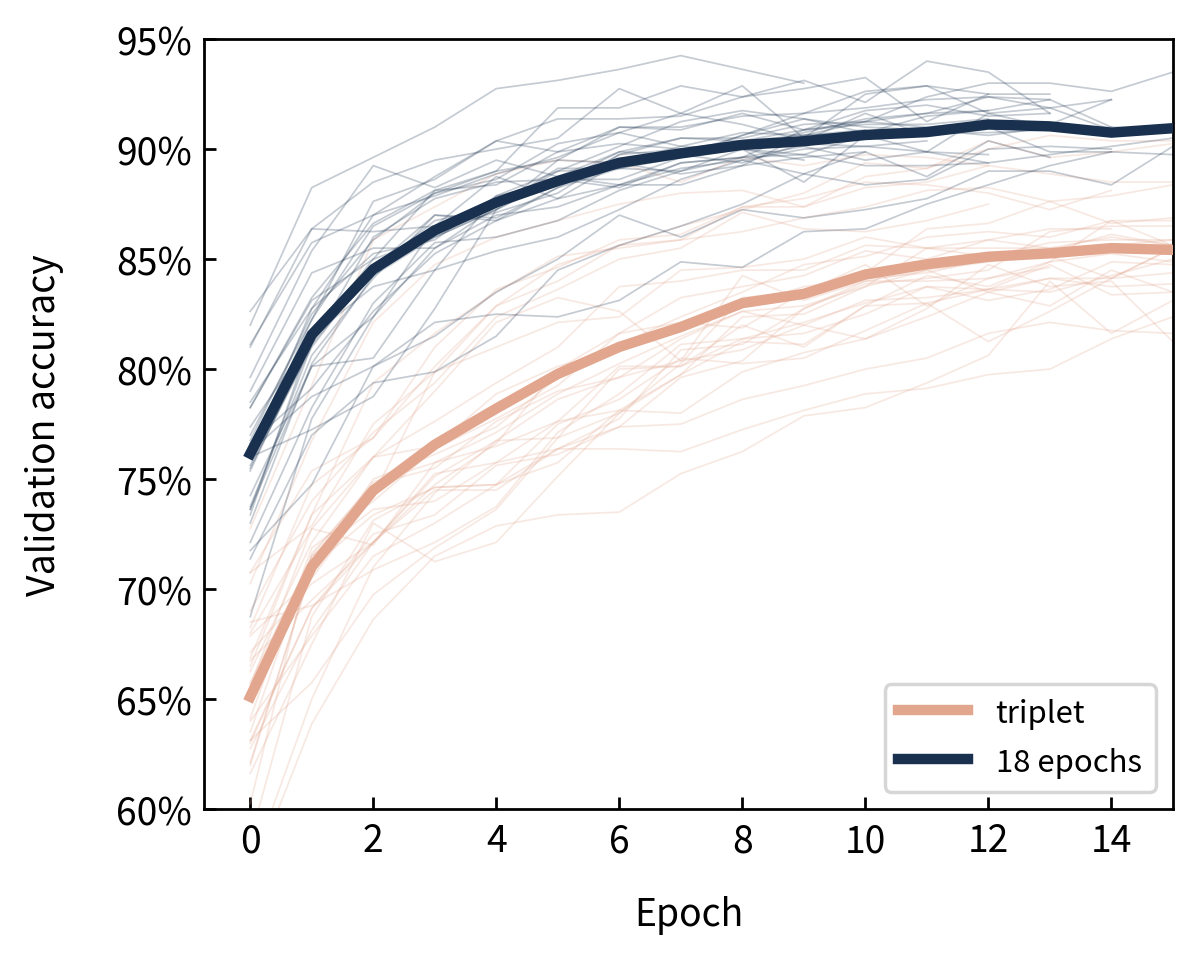}
    \caption{The validation accuracy for 30 iterations for each of the triplet model (pink) and 18 epoch model (blue), as evaluated on a validation set of 800 candidates. Thicker lines indicate the average accuracy for each model type.}
    \label{fig:model_comp}
\end{figure}

To quantify this boost in accuracy, we run 30 trials of hyperparameter sweeps for each model input type, fixing the architecture; parameter search range; and transient candidates in the training set. While these trials are a sparse, incomplete search for optimal hyperparameters, they reflect representative performance for each model input type. Figure \ref{fig:model_comp} shows the validation accuracy of both models, with the thicker lines indicating the average accuracy per model. We estimate a $\sim$5\% increase in validation accuracy when training on 18 epochs of difference imaging compared to triplet cutouts. We note that using time-dependent layers in our model (i.e. 3-dimensional convolutional layers and long short-term memory (LSTM) layers) to incorporate temporal information from our multi-epoch data offered no significant performance benefit over our adopted model.

\begin{figure}[h]
    \centering
    \includegraphics[width = \linewidth]{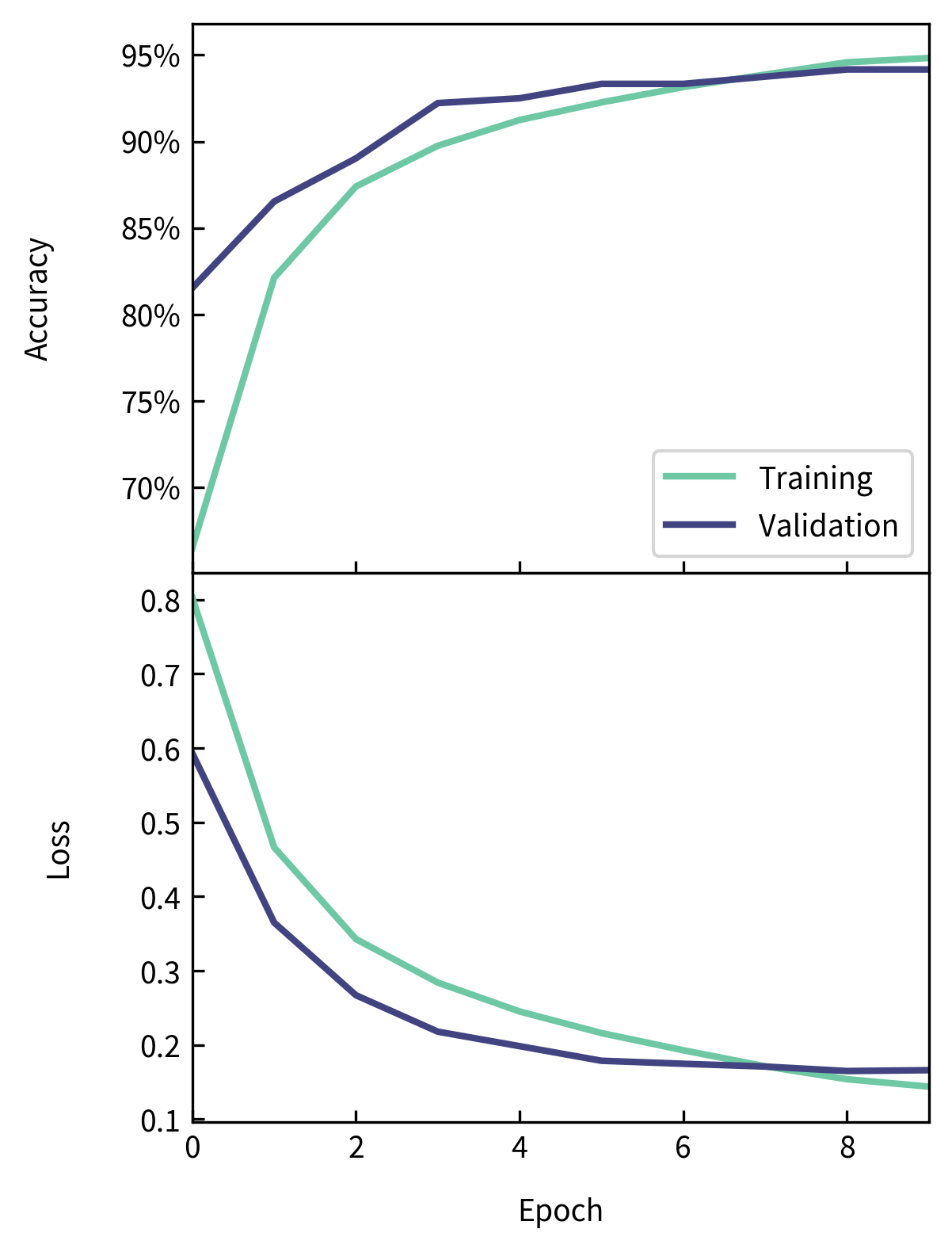}
    \caption{The accuracy (top) and loss (bottom) curves for the best performing model, computed on the training set (green) and validation set (blue). The model was trained for 10 epochs (iterations), and the early stopping criterion prevented the model from further training to avoid overfitting.}
    \label{fig:lossaccuracy}
\end{figure}

\begin{figure}[h]
    \centering
    \includegraphics[width = \linewidth]{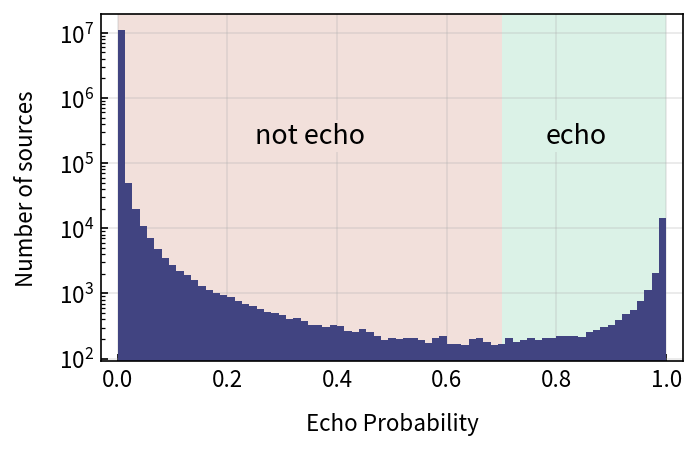}
    \caption{Histogram of the class probabilities of echoes for all transient candidates. The green shaded region shows candidates above our echo score threshold (shown in Figure \ref{fig:skydist}), while remaining candidates are in the red shaded region.}
    \label{fig:echohist}
\end{figure}

The adopted hyperparameters for our 18-epoch model are given in Table \ref{tab:multi_hps}, while the corresponding accuracy and loss curves are shown in Figure \ref{fig:lossaccuracy}.  
To further illustrate the overall performance of the echo model, we display the echo probabilities of all classified candidates from our echo search in Figure \ref{fig:echohist}. The bimodal distribution indicates that most sources have low probability of being classified as echoes, and the majority of echo candidates have high probabilities $>$ 0.95; therefore the classification accuracy is not very sensitive to the adopted echo score threshold of $> 0.7$.

\setlength{\arrayrulewidth}{0.2mm}

\begin{deluxetable}{c|cc}
\caption{Multi epoch model \label{tab:multi_hps}}
\tablehead{
    \colhead{Layer type} & \colhead{\makecell{Optimized \\ hyperparameter}} & \colhead{\makecell{Search range}}
}
\startdata
    2D conv. & 128 filters & 16\,--\,128 filters\\
    2D conv. & 128 filters & 16\,--\,128 filters\\
    2D conv. & 128 filters & 16\,--\,128 filters\\
    2D conv. & $3\times3$ kernel\tablenotemark{a} & - \\
    Max pool & $2\times2$ kernel\tablenotemark{a} & - \\
    Dropout & 0.4 & 0.1\,--\,0.5 \\
    Dense & 128 units & 16\,--\,128 units\\
    \hline
    \noalign{\vspace{3pt}}
    \multicolumn{3}{l}{Hyperparameter} \\
    \noalign{\vskip-\arrayrulewidth}
    \noalign{\vspace{3pt}}
    \hline
    Activation (conv.) & ReLU & [ReLU, Leaky ReLU] \\
    Activation (dense) & ReLU & [ReLU, Leaky ReLU] \\
    Learning Rate & $10^{-4}$ & $10^{-5}$-- $10^{-3}$ \\
    Batch Size & 32 & [16, 32, 64] \\
\enddata
\tablenotetext{a}{All convolutional and max pooling layers have zero padding. Stride is fixed at 1 for convolutional layers, and 2 for max pooling layers in accordance with the kernel size.}

\end{deluxetable}

\begin{deluxetable}{c|cc}
\vspace{6pt}
\caption{Single epoch model \label{tab:single_hps}}
\tablehead{
    \colhead{Layer type} & \colhead{\makecell{Optimized \\ hyperparameter}} & \colhead{\makecell{Search range}}
}
\startdata
    2D conv. & 32 filters & 16\,--\,128 filters\\
    2D conv. & 64 filters & 16\,--\,128 filters\\
    2D conv. & 128 filters & 16\,--\,128 filters\\
    2D conv. & $3\times3$ kernel\tablenotemark{a} & - \\
    Max pool & $2\times2$ kernel\tablenotemark{a} & - \\
    Dropout & 0.4 & 0.1\,--\,0.5 \\
    Dense & 128 units & 16\,--\,128 units\\
    \hline
    \noalign{\vspace{3pt}}
    \multicolumn{3}{l}{Hyperparameter} \\
    \noalign{\vskip-\arrayrulewidth}
    \noalign{\vspace{3pt}}
    \hline
    Activation (conv.) & ReLU & [ReLU, Leaky ReLU] \\
    Activation (dense) & Leaky ReLU & [ReLU, Leaky ReLU] \\
    Learning Rate & $10^{-4}$ & $10^{-5}$-- $10^{-3}$ \\
    Batch Size & 32 & [16, 32, 64] \\
\enddata
\tablenotetext{a}{All convolutional and max pooling layers have zero padding. Stride is fixed at 1 for convolutional layers, and 2 for max pooling layers in accordance with the kernel size.}
\end{deluxetable}



\section{Single epoch transient classification}
\label{sec:singleepoch}

\begin{figure*}[ht!]
    \includegraphics[width = \linewidth]{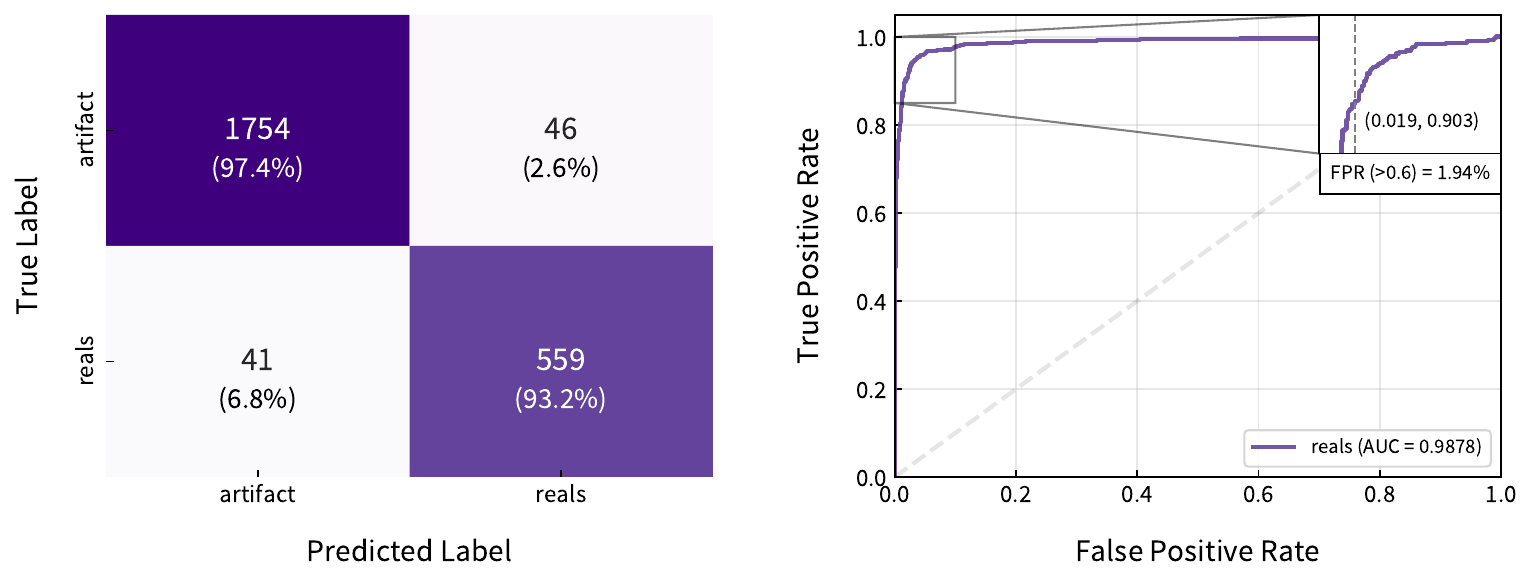}
    \caption{(Left) The confusion matrix for the best performing single epoch triplet model, as calculated using a test set of 2400 candidates. The confusion matrix is shown as a one-vs-all binary classification for reals, though the distribution of examples are balanced across classes. The percentages shown represent the fraction over the total number of sources for a particular class. (Right) The ROC curve for reals, characterized as a one-vs-all binary classification. We determine a threshold of 0.6 for true positives of reals at a FPR of 1.94\%. The inset shows a zoom-in of the top-left edge of the curve, used to define the score threshold.}
    \label{fig:confusion_triplet}
\end{figure*}

In this section, we briefly describe the model used on single epoch triplets to provide real/bogus scores for our catalog of NEOWISE transients. While real-bogus classification is typically established as a binary classification problem (e.g. \citealt{BTSbot, braii}), we found that preserving the multi-class setup for single epoch images performed better than binary classification for our training set. This holds true even when using techniques to mitigate class imbalance such as class re-weighting and down-sampling the number of examples in the majority class. We thus continue with a multi-classification model with the same four classes as the echo model, but prioritize the recovery of reals in our evaluation of model performance.

We employ an identical pre-processing procedure and model architecture as our original echo model, while only varying the input and the training process. Our training set of triplet examples includes 24,000 unique candidates, equally distributed between classes. We preserve the same 80\%/10\%/10\% split during model development, but do not include augmentations of the training data. We re-train the model using the triplet cutouts as our input, and conduct hyperparameter sweeps to determine the optimal hyperparameters for our model. The hyperparameters used for the final production model are shown at the bottom of Table \ref{tab:single_hps}.

We show the confusion matrix and ROC curve for our best-performing model in Figure \ref{fig:confusion_triplet}, adapting both as a real/bogus binary classification. From our test set of 2,400 candidates, the triplet model has an accuracy of 93.2\%, with a recall of 93.2\% and precision of 92.4\% for reals. From the ROC curve, we determine an AUC = 0.988 and a score of 0.6 for reals with a FPR of 1.94\%. This model is being employed for robust point source classification to aid transient and variable searches in the entire sample of NEOWISE candidates (K. De et al. in preparation), including candidates where the entire 18-epoch cutout set was not available due to the NEOWISE scanning strategy (e.g. incomplete coverage near the ecliptic poles, or towards the end of the mission).

\bibliography{ref}{}
\bibliographystyle{aasjournalv7}

\end{document}